\documentclass[aps,prl]{revtex4-1}

\usepackage{graphicx}
\usepackage{bm}
\usepackage{gensymb} 
\usepackage{color}
\usepackage{soul}
\usepackage{amsmath}
\usepackage{amssymb}

\begin{document}

\begin{center}

\Large{\textbf{\textit{Ab initio} molecular dynamics study of the structural and electronic transition in VO$_2$}}

\end{center}

\begin{center}

Du\v{s}an Pla\v{s}ienka$^\star$ and Roman Marto\v{n}\'{a}k

\textit{$^\star$ plasienka@fmph.uniba.sk}

\textit{Department of Experimental Physics, Comenius University in Bratislava, Bratislava, Slovakia}

\vspace{0.3cm}

Marcus C. Newton

\textit{Department of Physics \& Astronomy, University of Southampton, Southampton, UK}

\date{\today}

\end{center}


\textbf{Abstract:} The temperature-induced structural and electronic transformation in VO$_2$ between the monoclinic M1 and tetragonal rutile phases was studied by means of \textit{ab initio} molecular dynamics (MD), based on density functional theory with Hubbard correction (DFT+U). 
Analysis of the dynamical processes associated with the structural transformation was carried out on the atomic scale by following the time evolution of dimerization amplitudes of vanadium atom chains and the twisting angle of vanadium dimers. The electronic transition was studied by tracing the changes in projected densities of states and their correlation with the evolution of the  structural transformation. Our results reveal a strong interconnection between the structural and electronic transformations and show that they take place on the same time scale.



\section{Introduction}

\begin{figure}[h]
\includegraphics[width=0.9\columnwidth]{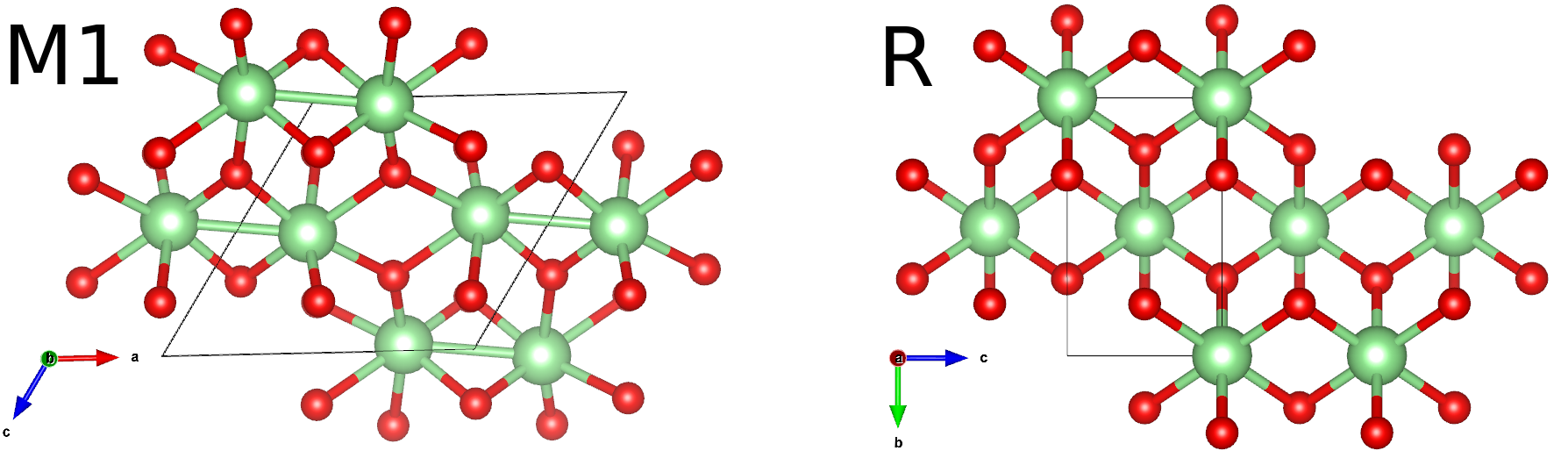}
\caption{Structure of M1 and R phases of VO$_2$ \cite{VESTA}.
Vanadium atoms are dimerized and octahedra are tilted in the monoclinic structure.}
\label{M1R}
\end{figure}

Vanadium dioxide VO$_2$ is a material of long standing interest and is one of the most  studied transition metal oxides \cite{Morin, Eyert-review, MIT-review, Wegkamp-Stahler}. At temperature of 340 K it exhibits a temperature-driven structural transition between low-temperature semiconducting non-magnetic monoclinic phase (M1) and high-temperature metallic paramagnetic tetragonal rutile
phase (R). The M1 phase has an optical gap of 0.6 eV. The insulator-metal transition is accompanied by dramatic change of
resistivity spanning over four orders of magnitude. The transition is first order and structurally represents a displacive transition.
In both M1 and R phases the V atoms are arranged in 1D chains. While in the R phase all V atoms are equidistant, in the M1 phase they
dimerize creating long bonds of length 3.18 \AA \,and short bonds of length 2.54 \AA - Fig.~\ref{M1R}. The dimerization is accompanied also by zig-zag deformation of chains and doubling of the unit cell in the chain direction. Other phases are also known, most notably another monoclinic phase M2 \cite{Park-VO2, Eyert} where only half of the chain is dimerized. This phase can be stabilized by doping or uniaxial stress but will not be addressed in the present study. 

While the transition in VO$_2$ is obviously of high interest for fundamental reasons, recently it has been speculated that it might
potentially also be of practical use in electronic switching and memristive devices \cite{Kim-switch, Yang, Driscoll, Kats}. The theoretical understanding of the transition, however, is still incomplete  and represents a puzzle despite a large number of works. The main problem  that was extensively studied is the very origin of the insulating properties  of the M1 phase. In this respect there are mainly two approaches. The first one interprets the transition within the standard band structure picture as being related to Peierls-type instability of the parent rutile phase. The work by Goodenough \cite{Goodenough, Eyert-review} suggested that the dimerization causes a splitting of the $d_{x^2-y^2}$ band while the zig-zag transversal displacements of V atoms cause a shift of the $d_{xz}$ and $d_{yz}$ bands. Both processes eventually result in creation of the band gap. The other direction attributes the metal-insulator transition upon cooling to strong correlations of d-electrons in V atoms resulting in the Mott insulator \cite{ZM}. Since the structural and electronic transitions appear to occur at the same time, it raises the ''chicken and egg'' question which of the two transitions is the
primary one. Both these questions and the role of the electronic correlations are still open and actively discussed in the literature
\cite{Wegkamp-Stahler, Cavalleri-3, Biermann, Baum, Cocker, Gray, Pashkin, Morrison, Wegkamp, Cocker, Laad, Koethe, Zhu, Gatti-1, Tomczak, Sakuma, Lazarovits, Wentzcovitch, Yao-VO2, Eyert, Brito, Zheng, Newton, Belozerov, Weber, Kim, Wall, Wall-2}.

Many studies were devoted to the possibilities of the description of the M1 and R phases by density functional theory (DFT), employing various
kinds of approximate functionals. The study in Ref. \cite{Wentzcovitch} showed that within the LDA approximation both phases are correctly
obtained as local minima of the Kohn-Sham energy. The M1 phase, however, resulted to be semi-metallic rather than insulating. Since 
both LDA and GGA approximations are known to underestimate the band gap,  in Refs. \cite{Budai, Kim} the DFT+U approach \cite{Anisimov-LDA+U, Liechtenstein} was employed with parameters U=4.2 eV and J=0.8 eV. For static calculations, it was found that this choice provides a satisfactory description of both M1 and R structures as well as a reasonable value of $\sim$0.6 eV for the band gap.

Considerably less attention was devoted to thermodynamical and dynamical aspects of the transition. Most studies focus on the ideal M1 and R phases at zero temperature and disregard the thermal fluctuations. It is known, however, since long time that the R phase has large value of the Debye-Waller factor which points to strong fluctuations of atomic positions \cite{Gervais, Budai, McWhan}. A recent study \cite{Budai} analyzed the phonon density of states in both phases, both experimentally and theoretically and concluded that the rutile phase is stabilized by entropy gain due to phonon softening in the metallic phase.

To our knowledge, so far there was no attempt to directly simulate the phase transition by means of \textit{ab initio} molecular dynamics. Our work attempts  to fill this gap in the literature by providing a detailed picture of the evolution of atomic and electronic structure across the thermally induced transition.

\section{Methods}

We simulated a fairly large system composed of 768 atoms by \textit{ab initio} MD in the NPT ensemble using the VASP code \cite{VASP-1, VASP-2, VASP-3, VASP-PAW}. In order to describe the system we chose to work within the PBE+U functional. This approximate functional appears as reasonable compromise between accuracy and computational cost. We found, however, that in order to get as close as possible to experimental conditions it is useful to modify the values of U and J. As noted in Ref. \cite{Budai} these values aside from the band gap affect also the energy difference between the M1 and R phases and therefore can be expected to directly influence also the transition temperature. For our purpose the most suitable values were found to be U=3.15 eV and J=0.6 eV. Further details about the simulation procedure and choice of U, J can be found in Suppl. Mat.

\section{Results}

\subsection{M1-R transition - atomic structure}

In order to monitor the evolution of the system across the transition we focus on two structural quantities - dimerization amplitudes (DA) and the zig-zag displacement (tilting of V-V dimers). Definitions of these two structural parameters are provided in Suppl. Mat.
We start with the transition induced by heating the M1 phase. At $T=400$ K we observed that the system attempts transitions from M1 to R phase but within 18 ps of the simulation time the monoclinic phase persisted. The evolution of dimerization amplitudes at this temperature is provided in Fig.~S3. It is quite possible that if we could wait a longer time the system could eventually transform from M1 to R. However, such a test would be prohibitively CPU time expensive. Increasing temperature to 450 K allowed the structural transition from M1 to R to complete within the reasonable CPU time of a few ps.  

\begin{figure}
\begin{tabular}{c|c}
\includegraphics[width=0.45\columnwidth]{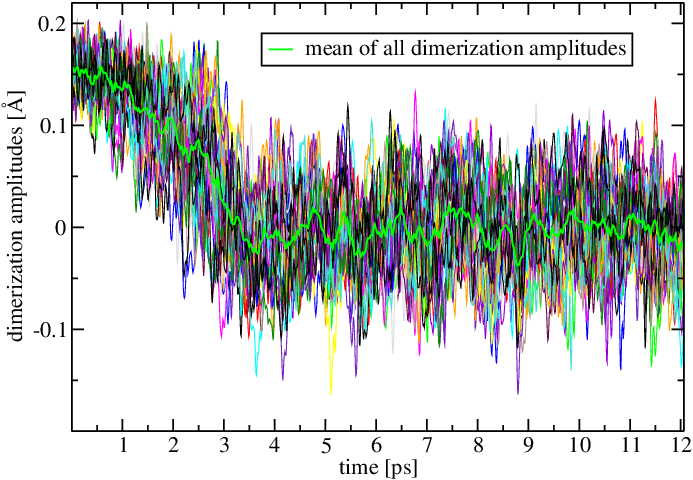}
\includegraphics[width=0.45\columnwidth]{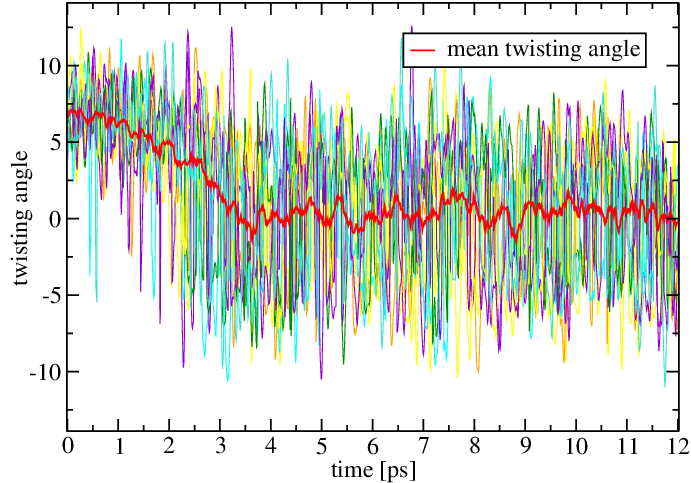}
\end{tabular}
\caption{M1 to R transition at 450 K showing sudden decrease of (a) dimerization amplitudes and of
(b) twisting angles. Different colors of curves distinguish between individual chain DA and dimer twisting angles,
while green curve in (a) and red curve in (b) represent mean DA and mean twisting angle, respectively.}
\label{M1-R}
\end{figure}

Fig.~\ref{M1-R} shows the evolution of DA and twisting angle across the transition. The decrease of DA from $\approx$ 0.15 \AA \,to 0 proceeds simultaneously with the adaptation of the twisting angle from 7$\degree$ to 0$\degree$ \cite{Yao-VO2}. This indicates no separation
between dimerization and tilting processes during the transition. After the transition 
to R, fluctuations of individual DA are considerably larger than in M1, which is directly
related to the large fluctuations of V atoms (see Suppl. Mat. for more details).

\begin{figure}
\includegraphics[width=0.7\columnwidth]{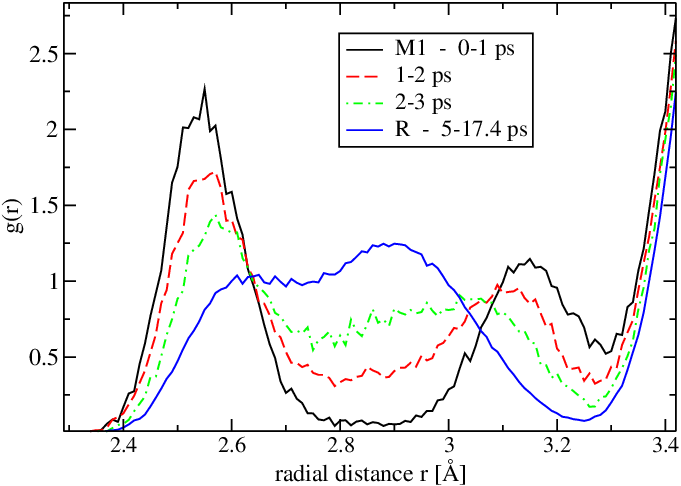}
\caption{Evolution of V-V RDF during the transition at 450 K.}
\label{RDF}
\end{figure}

Alongside with the evolution of DA and twisting angle, the evolution of the M1-R transformation can be
represented by plotting V-V radial distribution functions (RDFs) from several
short time intervals during the transformation. In Fig.~\ref{RDF}, there are four curves
showing V-V RDFs calculated from different time intervals - before the onset of the
transition in the M1 phase, during the transition and after it in the R phase. The M1
V-V RDF is characterized by relatively sharp two first peaks that upon transition merge into a broad
peak where the presence of two peaks is still visible. This shows that the fluctuations
of V atoms in the R phase above the transition are substantial.

\subsection{M1-R transition - electronic structure}

\begin{figure}
\includegraphics[width=0.5\columnwidth]{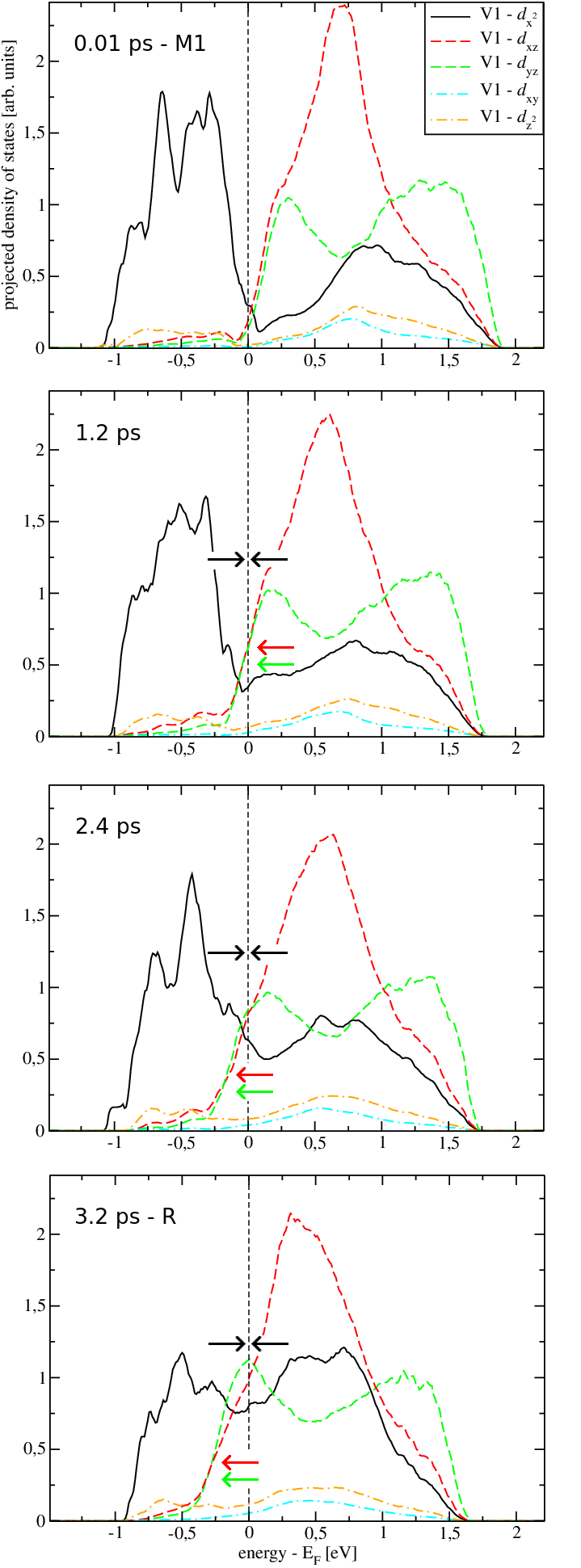}
\caption{Evolution of projected eDOS during the transition from initial M1 to final R phase at 450 K depicted at 0.01 ps (M1), 1.2 ps, 2.4 ps and at 3.2 ps (R).
Arrows denote principal changes observed for $d$-projected states.}
\label{eDOS}
\end{figure}

A standard band structure picture provides a description of M1 band gap as a mutual effect
of V-V dimerization, which causes splitting of V-$d_{x^2-y^2}$ states and zig-zag displacement
of dimers (tilting) that leads to energetic upshift of V-$d_{xz}$ and $d_{yz}$ states,
in the local geometry of V1 atoms \cite{Eyert-review, Eyert, Goodenough, Biermann, Lazarovits, Brito, Gray}.
The Peierls-like instability hence, in standard notation, applies to $a_{1g}$ states
in an embedded background of $e_g^{\pi}$ states.

The time evolution of projected electronic densities of states (peDOS) during the M1 to R
phase transition that occurred in the simulations at 450 K is shown in Fig.~\ref{eDOS}.
We note that within the applied PBE+U scheme the band gap drops to nearly zero value 
even before the transition, effectively turning the M1 phase, semiconducting at $T=0$, into semimetal.
The principal change in peDOS applies for $a_{1g}$ states that, after being split in dimerized M1 (0.01 ps),
become merged after the transition into R (3.2 ps). At the same time, disappearance of tilting
brings energy of $e^{\pi}_g$ states below the Fermi level E$_\textnormal{F}$ and the states
become partially occupied. The V1-$d_{z^2}$ and $d_{xy}$-derived states, which participate in
V-O bonds, do not change upon the transition much.

The correlation between electronic and atomic structure (Fig.~\ref{eDOS} and Fig.~\ref{M1-R}, respectively)
indicates that these two aspects of M1-R transformation are closely related and we did
not observe any evidence of separation between electronic and structural transition. Also, the two
different types of atomic motion - loss of dimerization and loss of tilting, that are directly
reflected in $a_1^g$ anti-splitting and $e^{\pi}_g$ energy lowering, respectively, proceed
mutually as well, as one could already conclude from Fig.~\ref{M1-R}.

\subsection{R-M1 transition}

\begin{figure}[h]
\includegraphics[width=0.7\columnwidth]{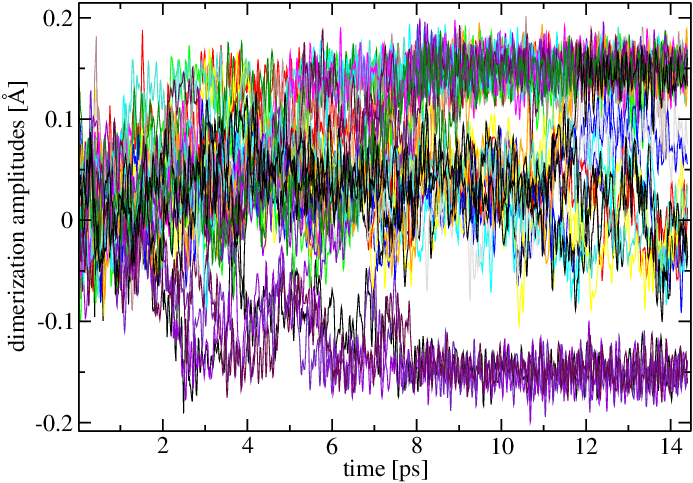}
\caption{Evolution of dimerization amplitudes during slow cooling of R phase to 300 K. Various colors differentiate between DA of individual chains.}
\label{cooling}
\end{figure}

After obtaining the R phase, we also tried to decrease the temperature
in order to study the symmetry-breaking process of the reverse R to M1 transition. 
We found that cooling rate (which is adjustable in NPT simulation) crucially
influences the final product of dimerization process. If fast quenching was
applied (temperature dropped by 100 K in 2 ps), a dimerized form with
randomly distributed V-V dimers was created. On the other hand, if considerably
slower cooling rate was applied (decrease by 100 K lasted for over 15 ps),
a much more organized state was obtained - Fig.~\ref{cooling}. In this state, most of the individual chains
emerged properly, but their overall arrangement was not regular as is in the M1 phase.

The failure of V-V dimers to organize on a long-range scale may be attributed
to the limited simulation time that is accessible in first-principles dynamical studies.
The system upon cooling dimerizes in order to decrease its energy, but follows a
randomly chosen pathway on the energy landscape, which usually leads into a metastable local
minimum (corresponding to dimerized, but not fully-organized crystalline state).
The system afterwards remains trapped in this state and reaching the global minimum requires a longer time scale. This might be partially avoided by using very slow thermal equilibration enabling system to explore a larger region of the configuration space, in analogy to the phenomenon of glass formation via rapid melt quenching, where disordered system is formed from the liquid upon fast quenching, while at lower cooling rates a regular crystal can be obtained instead.

\subsection{coherent X-ray diffraction comparison}

It is useful to contrast \textit{ab initio} calculations with complementary ultra-fast pump-probe coherent X-ray diffraction experiments that provide access to femto-second time-resolution and angstrom spatial resolution. Such experiments have become feasible with the recent advent of 4\textsuperscript{th} generation X-ray free electron laser (XFEL) synchrotron facilities such as the European XFEL. The spatially varying far-field scattering amplitude distribution is proportional to the exponentiated phase associated with scattering from each atom in the supercell. Monitoring changes in the simulated scattering angle and intensity from the speckle pattern provides direct comparison of theory with experiment (see Suppl. Mat.). 

\begin{figure}[h]
\includegraphics[width=0.7\columnwidth]{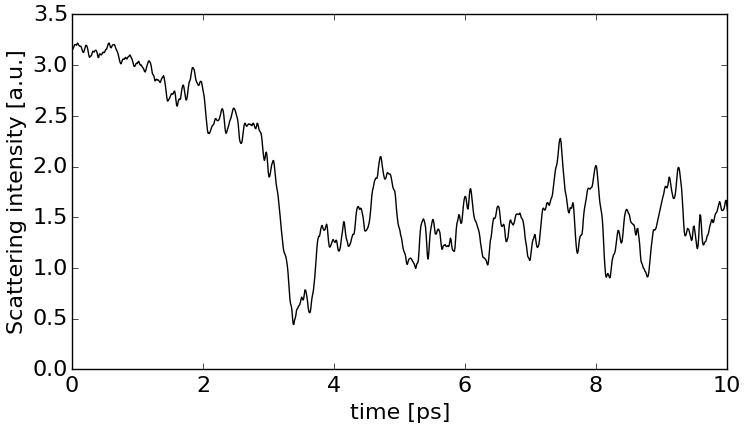}
\caption{Scattering intensity of centroid position during simulation at 450 K. A dip in intensity coincides with the onset of the structural transition.}
\label{centroid}
\end{figure}

Fig.~\ref{centroid} shows the intensity of the resulting speckle patterns' centroid plotted as a function of $t$. As the total intensity of the speckle pattern is dependent on the amplitude of the complex object, only the relative value of the diffraction patterns' intensity is meaningful in the simulations. A characteristic dip lasting for 400 fs is observed immediately before the rutile phase begins at ca. 3.5 ps, which corresponds nicely with the onset of dimerization (Fig.~\ref{M1-R}).  This is also observed experimentally (See Newton \textit{et al.}, Ref. \cite{Newton}) and can be attributed to destructive interference due to disordering at the onset of the structural transition. The intensity dip is also understood from the perspective of fundamental Fourier optics where it can be shown that a phase object imaged with a limited transfer function produces sharp reduction in scattering intensity precisely at the location of the phase change. \cite{Goodman}

\section{Conclusions}

In summary, we studied the structural and electronic transition in VO$_2$ by means of \textit{ab initio} molecular dynamics in variable-cell approach. Within the employed PBE+U description, the temperature-induced transition from M1 to R phase was observed at the transition temperature of 450 K. We found that dimerization and tilting of octahedra evolve at the same time scale and electronic and structural changes occur concurrently, as revealed by detailed microscopic analysis of structural and electronic transformations. The reverse process of dimerization upon cooling the rutile phase was found, however, to be critically contingent upon the rate of cooling. Moreover, we find that the high-temperature rutile phase is characterized by large fluctuations of positions of V atoms, in good agreement with the large value of the Debye-Waller factor known from experiment.

\section{acknowledgments}

This work was supported by the Slovak Research and Development Agency under Contract APVV-15-0496 and
VEGA project No. 1/0904/15 and by the project implementation 26220220004 within the Research \& Development
Operational Programme funded by the ERDF. This work is funded in part by Royal Society Research Grant RG/130498.
The authors acknowledge the use of the IRIDIS High Performance Computing Facility, and associated support
services at the University of Southampton, in the completion of this work. Part of the calculations were
also performed in the Computing Centre of the Slovak Academy of Sciences using the supercomputing infrastructure
acquired in project ITMS 26230120002 and 26210120002 (Slovak infrastructure for high-performance computing)
supported by the Research \& Development Operational Programme funded by the ERDF.



\begin{thebibliography}{54}
\expandafter\ifx\csname natexlab\endcsname\relax\def\natexlab#1{#1}\fi
\expandafter\ifx\csname bibnamefont\endcsname\relax
  \def\bibnamefont#1{#1}\fi
\expandafter\ifx\csname bibfnamefont\endcsname\relax
  \def\bibfnamefont#1{#1}\fi
\expandafter\ifx\csname citenamefont\endcsname\relax
  \def\citenamefont#1{#1}\fi
\expandafter\ifx\csname url\endcsname\relax
  \def\url#1{\texttt{#1}}\fi
\expandafter\ifx\csname urlprefix\endcsname\relax\def\urlprefix{URL }\fi
\providecommand{\bibinfo}[2]{#2}
\providecommand{\eprint}[2][]{\url{#2}}

\bibitem[{\citenamefont{Morin}(1959)}]{Morin}
\bibinfo{author}{\bibfnamefont{F.~J.} \bibnamefont{Morin}},
  \bibinfo{journal}{Phys. Rev. Lett.} \textbf{\bibinfo{volume}{3}},
  \bibinfo{pages}{34} (\bibinfo{year}{1959}).

\bibitem[{\citenamefont{Eyert}(2002)}]{Eyert-review}
\bibinfo{author}{\bibfnamefont{V.}~\bibnamefont{Eyert}}, \bibinfo{journal}{Ann.
  Phys. (Leipzig)} \textbf{\bibinfo{volume}{11}}, \bibinfo{pages}{650}
  (\bibinfo{year}{2002}).

\bibitem[{\citenamefont{Imada et~al.}(1998)\citenamefont{Imada, Fujimori, and
  Tokura}}]{MIT-review}
\bibinfo{author}{\bibfnamefont{M.}~\bibnamefont{Imada}},
  \bibinfo{author}{\bibfnamefont{A.}~\bibnamefont{Fujimori}}, \bibnamefont{and}
  \bibinfo{author}{\bibfnamefont{Y.}~\bibnamefont{Tokura}},
  \bibinfo{journal}{Rev. Mod. Phys.} \textbf{\bibinfo{volume}{70}},
  \bibinfo{pages}{1039} (\bibinfo{year}{1998}).

\bibitem[{\citenamefont{Wegkamp and St\"{a}hler}(2015)}]{Wegkamp-Stahler}
\bibinfo{author}{\bibfnamefont{D.}~\bibnamefont{Wegkamp}} \bibnamefont{and}
  \bibinfo{author}{\bibfnamefont{J.}~\bibnamefont{St\"{a}hler}},
  \bibinfo{journal}{Prog. Surf. Sci.} \textbf{\bibinfo{volume}{90}},
  \bibinfo{pages}{464} (\bibinfo{year}{2015}).

\bibitem[{\citenamefont{Park et~al.}(2013)\citenamefont{Park, Coy, Kasirga,
  Huang, Fei, Hunter, and Cobden}}]{Park-VO2}
\bibinfo{author}{\bibfnamefont{J.~H.} \bibnamefont{Park}},
  \bibinfo{author}{\bibfnamefont{J.~M.} \bibnamefont{Coy}},
  \bibinfo{author}{\bibfnamefont{T.~S.} \bibnamefont{Kasirga}},
  \bibinfo{author}{\bibfnamefont{C.}~\bibnamefont{Huang}},
  \bibinfo{author}{\bibfnamefont{Z.}~\bibnamefont{Fei}},
  \bibinfo{author}{\bibfnamefont{S.}~\bibnamefont{Hunter}}, \bibnamefont{and}
  \bibinfo{author}{\bibfnamefont{D.~H.} \bibnamefont{Cobden}},
  \bibinfo{journal}{Nature} \textbf{\bibinfo{volume}{500}},
  \bibinfo{pages}{431} (\bibinfo{year}{2013}).

\bibitem[{\citenamefont{Eyert}(2011)}]{Eyert}
\bibinfo{author}{\bibfnamefont{V.}~\bibnamefont{Eyert}},
  \bibinfo{journal}{Phys. Rev. Lett.} \textbf{\bibinfo{volume}{107}},
  \bibinfo{pages}{016401} (\bibinfo{year}{2011}).

\bibitem[{\citenamefont{Kim et~al.}(2010)\citenamefont{Kim, Ko, Frenzel,
  Ramanathan, and Hoffman}}]{Kim-switch}
\bibinfo{author}{\bibfnamefont{J.}~\bibnamefont{Kim}},
  \bibinfo{author}{\bibfnamefont{C.}~\bibnamefont{Ko}},
  \bibinfo{author}{\bibfnamefont{A.}~\bibnamefont{Frenzel}},
  \bibinfo{author}{\bibfnamefont{S.}~\bibnamefont{Ramanathan}},
  \bibnamefont{and} \bibinfo{author}{\bibfnamefont{J.~E.}
  \bibnamefont{Hoffman}}, \bibinfo{journal}{Appl. Phys. Lett.}
  \textbf{\bibinfo{volume}{96}}, \bibinfo{pages}{213106}
  (\bibinfo{year}{2010}).

\bibitem[{\citenamefont{Yang et~al.}(2011)\citenamefont{Yang, Ko, and
  Ramanathan}}]{Yang}
\bibinfo{author}{\bibfnamefont{Z.}~\bibnamefont{Yang}},
  \bibinfo{author}{\bibfnamefont{C.}~\bibnamefont{Ko}}, \bibnamefont{and}
  \bibinfo{author}{\bibfnamefont{S.}~\bibnamefont{Ramanathan}},
  \bibinfo{journal}{Annu. Rev. Mater. Res.} \textbf{\bibinfo{volume}{41}},
  \bibinfo{pages}{337} (\bibinfo{year}{2011}).

\bibitem[{\citenamefont{Driscoll et~al.}(2009)\citenamefont{Driscoll, Kim,
  Chae, Ventra, and Basov}}]{Driscoll}
\bibinfo{author}{\bibfnamefont{T.}~\bibnamefont{Driscoll}},
  \bibinfo{author}{\bibfnamefont{H.-T.} \bibnamefont{Kim}},
  \bibinfo{author}{\bibfnamefont{B.-G.} \bibnamefont{Chae}},
  \bibinfo{author}{\bibfnamefont{M.~D.} \bibnamefont{Ventra}},
  \bibnamefont{and} \bibinfo{author}{\bibfnamefont{D.~N.} \bibnamefont{Basov}},
  \bibinfo{journal}{Appl. Phys. Lett.} \textbf{\bibinfo{volume}{95}},
  \bibinfo{pages}{043503} (\bibinfo{year}{2009}).

\bibitem[{\citenamefont{Kats et~al.}(2013)\citenamefont{Kats, Blanchard, Zhang,
  Genevet, Ko, Ramanathan, and Capasso}}]{Kats}
\bibinfo{author}{\bibfnamefont{M.~A.} \bibnamefont{Kats}},
  \bibinfo{author}{\bibfnamefont{R.}~\bibnamefont{Blanchard}},
  \bibinfo{author}{\bibfnamefont{S.}~\bibnamefont{Zhang}},
  \bibinfo{author}{\bibfnamefont{P.}~\bibnamefont{Genevet}},
  \bibinfo{author}{\bibfnamefont{C.}~\bibnamefont{Ko}},
  \bibinfo{author}{\bibfnamefont{S.}~\bibnamefont{Ramanathan}},
  \bibnamefont{and} \bibinfo{author}{\bibfnamefont{F.}~\bibnamefont{Capasso}},
  \bibinfo{journal}{Phys. Rev. X} \textbf{\bibinfo{volume}{3}},
  \bibinfo{pages}{041004} (\bibinfo{year}{2013}).

\bibitem[{\citenamefont{Goodenough}(1971)}]{Goodenough}
\bibinfo{author}{\bibfnamefont{J.~B.} \bibnamefont{Goodenough}},
  \bibinfo{journal}{J. Solid Stat. Chem.} \textbf{\bibinfo{volume}{3}},
  \bibinfo{pages}{490} (\bibinfo{year}{1971}).

\bibitem[{\citenamefont{Zylbersztejn and Mott}(1975)}]{ZM}
\bibinfo{author}{\bibfnamefont{A.}~\bibnamefont{Zylbersztejn}}
  \bibnamefont{and} \bibinfo{author}{\bibfnamefont{N.~F.} \bibnamefont{Mott}},
  \bibinfo{journal}{Phys. Rev. B} \textbf{\bibinfo{volume}{11}},
  \bibinfo{pages}{4383} (\bibinfo{year}{1975}).

\bibitem[{\citenamefont{Cavalleri et~al.}(2005)\citenamefont{Cavalleri, Rini,
  Chong, Fourmaux, Glover, Heimann, Kieffer, and Schoenlein}}]{Cavalleri-3}
\bibinfo{author}{\bibfnamefont{A.}~\bibnamefont{Cavalleri}},
  \bibinfo{author}{\bibfnamefont{M.}~\bibnamefont{Rini}},
  \bibinfo{author}{\bibfnamefont{H.~H.~W.} \bibnamefont{Chong}},
  \bibinfo{author}{\bibfnamefont{S.}~\bibnamefont{Fourmaux}},
  \bibinfo{author}{\bibfnamefont{T.~E.} \bibnamefont{Glover}},
  \bibinfo{author}{\bibfnamefont{P.~A.} \bibnamefont{Heimann}},
  \bibinfo{author}{\bibfnamefont{J.~C.} \bibnamefont{Kieffer}},
  \bibnamefont{and} \bibinfo{author}{\bibfnamefont{R.~W.}
  \bibnamefont{Schoenlein}}, \bibinfo{journal}{Phys. Rev. Lett.}
  \textbf{\bibinfo{volume}{95}}, \bibinfo{pages}{067405}
  (\bibinfo{year}{2005}).

\bibitem[{\citenamefont{Biermann et~al.}(2005)\citenamefont{Biermann,
  Poteryaev, Lichtenstein, and Georges}}]{Biermann}
\bibinfo{author}{\bibfnamefont{S.}~\bibnamefont{Biermann}},
  \bibinfo{author}{\bibfnamefont{A.}~\bibnamefont{Poteryaev}},
  \bibinfo{author}{\bibfnamefont{A.~I.} \bibnamefont{Lichtenstein}},
  \bibnamefont{and} \bibinfo{author}{\bibfnamefont{A.}~\bibnamefont{Georges}},
  \bibinfo{journal}{Phys. Rev. Lett.} \textbf{\bibinfo{volume}{94}},
  \bibinfo{pages}{026404} (\bibinfo{year}{2005}).

\bibitem[{\citenamefont{Baum et~al.}(2007)\citenamefont{Baum, Yang, and
  Zewail}}]{Baum}
\bibinfo{author}{\bibfnamefont{P.}~\bibnamefont{Baum}},
  \bibinfo{author}{\bibfnamefont{D.-S.} \bibnamefont{Yang}}, \bibnamefont{and}
  \bibinfo{author}{\bibfnamefont{A.~H.} \bibnamefont{Zewail}},
  \bibinfo{journal}{Science} \textbf{\bibinfo{volume}{318}},
  \bibinfo{pages}{788} (\bibinfo{year}{2007}).

\bibitem[{\citenamefont{Cocker et~al.}(2012)\citenamefont{Cocker, Titova,
  Fourmaux, Holloway, Bandulet, Brassard, Kieffer, ElKhakani, and
  Hegmann}}]{Cocker}
\bibinfo{author}{\bibfnamefont{T.~L.} \bibnamefont{Cocker}},
  \bibinfo{author}{\bibfnamefont{L.~V.} \bibnamefont{Titova}},
  \bibinfo{author}{\bibfnamefont{S.}~\bibnamefont{Fourmaux}},
  \bibinfo{author}{\bibfnamefont{G.}~\bibnamefont{Holloway}},
  \bibinfo{author}{\bibfnamefont{H.-C.} \bibnamefont{Bandulet}},
  \bibinfo{author}{\bibfnamefont{D.}~\bibnamefont{Brassard}},
  \bibinfo{author}{\bibfnamefont{J.-C.} \bibnamefont{Kieffer}},
  \bibinfo{author}{\bibfnamefont{M.~A.} \bibnamefont{ElKhakani}},
  \bibnamefont{and} \bibinfo{author}{\bibfnamefont{F.~A.}
  \bibnamefont{Hegmann}}, \bibinfo{journal}{Phys. Rev. B}
  \textbf{\bibinfo{volume}{85}}, \bibinfo{pages}{155120}
  (\bibinfo{year}{2012}).

\bibitem[{\citenamefont{Gray et~al.}(2016)\citenamefont{Gray, Jeong, Aetukuri,
  Granitzka, Chen, Kukreja, Higley, Chase, Reid, Ohldag et~al.}}]{Gray}
\bibinfo{author}{\bibfnamefont{A.~X.} \bibnamefont{Gray}},
  \bibinfo{author}{\bibfnamefont{J.}~\bibnamefont{Jeong}},
  \bibinfo{author}{\bibfnamefont{N.~P.} \bibnamefont{Aetukuri}},
  \bibinfo{author}{\bibfnamefont{P.}~\bibnamefont{Granitzka}},
  \bibinfo{author}{\bibfnamefont{Z.}~\bibnamefont{Chen}},
  \bibinfo{author}{\bibfnamefont{R.}~\bibnamefont{Kukreja}},
  \bibinfo{author}{\bibfnamefont{D.}~\bibnamefont{Higley}},
  \bibinfo{author}{\bibfnamefont{T.}~\bibnamefont{Chase}},
  \bibinfo{author}{\bibfnamefont{A.~H.} \bibnamefont{Reid}},
  \bibinfo{author}{\bibfnamefont{H.}~\bibnamefont{Ohldag}},
  \bibnamefont{et~al.}, \bibinfo{journal}{Phys. Rev. Lett.}
  \textbf{\bibinfo{volume}{116}}, \bibinfo{pages}{116403}
  (\bibinfo{year}{2016}).

\bibitem[{\citenamefont{Pashkin et~al.}(2011)\citenamefont{Pashkin, K\"{u}bler,
  Ehrke, Lopez, Halabica, Haglund, Huber, and Leitenstorfer}}]{Pashkin}
\bibinfo{author}{\bibfnamefont{A.}~\bibnamefont{Pashkin}},
  \bibinfo{author}{\bibfnamefont{C.}~\bibnamefont{K\"{u}bler}},
  \bibinfo{author}{\bibfnamefont{H.}~\bibnamefont{Ehrke}},
  \bibinfo{author}{\bibfnamefont{R.}~\bibnamefont{Lopez}},
  \bibinfo{author}{\bibfnamefont{A.}~\bibnamefont{Halabica}},
  \bibinfo{author}{\bibfnamefont{R.~F.} \bibnamefont{Haglund}},
  \bibinfo{author}{\bibfnamefont{R.}~\bibnamefont{Huber}}, \bibnamefont{and}
  \bibinfo{author}{\bibfnamefont{A.}~\bibnamefont{Leitenstorfer}},
  \bibinfo{journal}{Phys. Rev. B} \textbf{\bibinfo{volume}{83}},
  \bibinfo{pages}{195120} (\bibinfo{year}{2011}).

\bibitem[{\citenamefont{Morrison et~al.}(2014)\citenamefont{Morrison,
  Chatelain, Tiwari, Hendaoui, Bruh\'{a}cs, Chaker, and Siwick}}]{Morrison}
\bibinfo{author}{\bibfnamefont{V.~R.} \bibnamefont{Morrison}},
  \bibinfo{author}{\bibfnamefont{R.~P.} \bibnamefont{Chatelain}},
  \bibinfo{author}{\bibfnamefont{K.~L.} \bibnamefont{Tiwari}},
  \bibinfo{author}{\bibfnamefont{A.}~\bibnamefont{Hendaoui}},
  \bibinfo{author}{\bibfnamefont{A.}~\bibnamefont{Bruh\'{a}cs}},
  \bibinfo{author}{\bibfnamefont{M.}~\bibnamefont{Chaker}}, \bibnamefont{and}
  \bibinfo{author}{\bibfnamefont{B.~J.} \bibnamefont{Siwick}},
  \bibinfo{journal}{Science} \textbf{\bibinfo{volume}{346}},
  \bibinfo{pages}{445} (\bibinfo{year}{2014}).

\bibitem[{\citenamefont{Wegkamp et~al.}(2014)\citenamefont{Wegkamp, Herzog,
  Xian, Gatti, Cudazzo, McGahan, Marvel, Richard F.~Haglund, Rubio, Wolf
  et~al.}}]{Wegkamp}
\bibinfo{author}{\bibfnamefont{D.}~\bibnamefont{Wegkamp}},
  \bibinfo{author}{\bibfnamefont{M.}~\bibnamefont{Herzog}},
  \bibinfo{author}{\bibfnamefont{L.}~\bibnamefont{Xian}},
  \bibinfo{author}{\bibfnamefont{M.}~\bibnamefont{Gatti}},
  \bibinfo{author}{\bibfnamefont{P.}~\bibnamefont{Cudazzo}},
  \bibinfo{author}{\bibfnamefont{C.~L.} \bibnamefont{McGahan}},
  \bibinfo{author}{\bibfnamefont{R.~E.} \bibnamefont{Marvel}},
  \bibinfo{author}{\bibfnamefont{J.}~\bibnamefont{Richard F.~Haglund}},
  \bibinfo{author}{\bibfnamefont{A.}~\bibnamefont{Rubio}},
  \bibinfo{author}{\bibfnamefont{M.}~\bibnamefont{Wolf}}, \bibnamefont{et~al.},
  \bibinfo{journal}{Phys. Rev. Lett.} \textbf{\bibinfo{volume}{113}},
  \bibinfo{pages}{216401} (\bibinfo{year}{2014}).

\bibitem[{\citenamefont{Laad et~al.}(2006)\citenamefont{Laad, Craco, and
  M\"{u}ller-Hartmann}}]{Laad}
\bibinfo{author}{\bibfnamefont{M.~S.} \bibnamefont{Laad}},
  \bibinfo{author}{\bibfnamefont{L.}~\bibnamefont{Craco}}, \bibnamefont{and}
  \bibinfo{author}{\bibfnamefont{E.}~\bibnamefont{M\"{u}ller-Hartmann}},
  \bibinfo{journal}{Phys. Rev. B} \textbf{\bibinfo{volume}{73}},
  \bibinfo{pages}{195120} (\bibinfo{year}{2006}).

\bibitem[{\citenamefont{Koethe et~al.}(2006)\citenamefont{Koethe, Hu,
  Haverkort, Schussler-Langeheine, Venturini, Brookes, Tjernberg, Reichelt,
  Hsieh, Lin et~al.}}]{Koethe}
\bibinfo{author}{\bibfnamefont{T.~C.} \bibnamefont{Koethe}},
  \bibinfo{author}{\bibfnamefont{Z.}~\bibnamefont{Hu}},
  \bibinfo{author}{\bibfnamefont{M.~W.} \bibnamefont{Haverkort}},
  \bibinfo{author}{\bibfnamefont{C.}~\bibnamefont{Schussler-Langeheine}},
  \bibinfo{author}{\bibfnamefont{F.}~\bibnamefont{Venturini}},
  \bibinfo{author}{\bibfnamefont{N.~B.} \bibnamefont{Brookes}},
  \bibinfo{author}{\bibfnamefont{O.}~\bibnamefont{Tjernberg}},
  \bibinfo{author}{\bibfnamefont{W.}~\bibnamefont{Reichelt}},
  \bibinfo{author}{\bibfnamefont{H.~H.} \bibnamefont{Hsieh}},
  \bibinfo{author}{\bibfnamefont{H.-J.} \bibnamefont{Lin}},
  \bibnamefont{et~al.}, \bibinfo{journal}{Phys. Rev. Lett.}
  \textbf{\bibinfo{volume}{97}}, \bibinfo{pages}{116402}
  (\bibinfo{year}{2006}).

\bibitem[{\citenamefont{Zhu and Schwingenschl\"{o}gl}(2012)}]{Zhu}
\bibinfo{author}{\bibfnamefont{Z.}~\bibnamefont{Zhu}} \bibnamefont{and}
  \bibinfo{author}{\bibfnamefont{U.}~\bibnamefont{Schwingenschl\"{o}gl}},
  \bibinfo{journal}{Phys. Rev. B} \textbf{\bibinfo{volume}{86}},
  \bibinfo{pages}{075149} (\bibinfo{year}{2012}).

\bibitem[{\citenamefont{Gatti et~al.}(2007)\citenamefont{Gatti, Bruneval,
  Olevano, and Reining}}]{Gatti-1}
\bibinfo{author}{\bibfnamefont{M.}~\bibnamefont{Gatti}},
  \bibinfo{author}{\bibfnamefont{F.}~\bibnamefont{Bruneval}},
  \bibinfo{author}{\bibfnamefont{V.}~\bibnamefont{Olevano}}, \bibnamefont{and}
  \bibinfo{author}{\bibfnamefont{L.}~\bibnamefont{Reining}},
  \bibinfo{journal}{Phys. Rev. Lett.} \textbf{\bibinfo{volume}{99}},
  \bibinfo{pages}{266402} (\bibinfo{year}{2007}).

\bibitem[{\citenamefont{Tomczak et~al.}(2008)\citenamefont{Tomczak,
  Aryasetiawan, and Biermann}}]{Tomczak}
\bibinfo{author}{\bibfnamefont{J.~M.} \bibnamefont{Tomczak}},
  \bibinfo{author}{\bibfnamefont{F.}~\bibnamefont{Aryasetiawan}},
  \bibnamefont{and} \bibinfo{author}{\bibfnamefont{S.}~\bibnamefont{Biermann}},
  \bibinfo{journal}{Phys. Rev. B} \textbf{\bibinfo{volume}{78}},
  \bibinfo{pages}{115103} (\bibinfo{year}{2008}).

\bibitem[{\citenamefont{Sakuma et~al.}(2008)\citenamefont{Sakuma, Miyake, and
  Aryasetiawan}}]{Sakuma}
\bibinfo{author}{\bibfnamefont{R.}~\bibnamefont{Sakuma}},
  \bibinfo{author}{\bibfnamefont{T.}~\bibnamefont{Miyake}}, \bibnamefont{and}
  \bibinfo{author}{\bibfnamefont{F.}~\bibnamefont{Aryasetiawan}},
  \bibinfo{journal}{Phys. Rev. B} \textbf{\bibinfo{volume}{78}},
  \bibinfo{pages}{075106} (\bibinfo{year}{2008}).

\bibitem[{\citenamefont{Lazarovits et~al.}(2010)\citenamefont{Lazarovits, Kim,
  Haule, and Kotliar}}]{Lazarovits}
\bibinfo{author}{\bibfnamefont{B.}~\bibnamefont{Lazarovits}},
  \bibinfo{author}{\bibfnamefont{K.}~\bibnamefont{Kim}},
  \bibinfo{author}{\bibfnamefont{K.}~\bibnamefont{Haule}}, \bibnamefont{and}
  \bibinfo{author}{\bibfnamefont{G.}~\bibnamefont{Kotliar}},
  \bibinfo{journal}{Phys. Rev. B} \textbf{\bibinfo{volume}{81}},
  \bibinfo{pages}{115117} (\bibinfo{year}{2010}).

\bibitem[{\citenamefont{Wentzcovitch et~al.}(1994)\citenamefont{Wentzcovitch,
  Schulz, and Allen}}]{Wentzcovitch}
\bibinfo{author}{\bibfnamefont{R.~M.} \bibnamefont{Wentzcovitch}},
  \bibinfo{author}{\bibfnamefont{W.~W.} \bibnamefont{Schulz}},
  \bibnamefont{and} \bibinfo{author}{\bibfnamefont{P.~B.} \bibnamefont{Allen}},
  \bibinfo{journal}{Phys. Rev. Lett.} \textbf{\bibinfo{volume}{72}},
  \bibinfo{pages}{3389} (\bibinfo{year}{1994}).

\bibitem[{\citenamefont{Yao et~al.}(2010)\citenamefont{Yao, Zhang, Sun, Liu,
  Huang, Xie, Wu, Yuan, Zhang, Wu et~al.}}]{Yao-VO2}
\bibinfo{author}{\bibfnamefont{T.}~\bibnamefont{Yao}},
  \bibinfo{author}{\bibfnamefont{X.}~\bibnamefont{Zhang}},
  \bibinfo{author}{\bibfnamefont{Z.}~\bibnamefont{Sun}},
  \bibinfo{author}{\bibfnamefont{S.}~\bibnamefont{Liu}},
  \bibinfo{author}{\bibfnamefont{Y.}~\bibnamefont{Huang}},
  \bibinfo{author}{\bibfnamefont{Y.}~\bibnamefont{Xie}},
  \bibinfo{author}{\bibfnamefont{C.}~\bibnamefont{Wu}},
  \bibinfo{author}{\bibfnamefont{X.}~\bibnamefont{Yuan}},
  \bibinfo{author}{\bibfnamefont{W.}~\bibnamefont{Zhang}},
  \bibinfo{author}{\bibfnamefont{Z.}~\bibnamefont{Wu}}, \bibnamefont{et~al.},
  \bibinfo{journal}{Phys. Rev. Lett.} \textbf{\bibinfo{volume}{105}},
  \bibinfo{pages}{226405} (\bibinfo{year}{2010}).

\bibitem[{\citenamefont{Brito et~al.}(2016)\citenamefont{Brito, Aguiar, Haule,
  and Kotliar}}]{Brito}
\bibinfo{author}{\bibfnamefont{W.~H.} \bibnamefont{Brito}},
  \bibinfo{author}{\bibfnamefont{M.~C.~O.} \bibnamefont{Aguiar}},
  \bibinfo{author}{\bibfnamefont{K.}~\bibnamefont{Haule}}, \bibnamefont{and}
  \bibinfo{author}{\bibfnamefont{G.}~\bibnamefont{Kotliar}},
  \bibinfo{journal}{Phys. Rev. Lett.} \textbf{\bibinfo{volume}{117}},
  \bibinfo{pages}{056402} (\bibinfo{year}{2016}).

\bibitem[{\citenamefont{Zheng and Wagner}(2015)}]{Zheng}
\bibinfo{author}{\bibfnamefont{H.}~\bibnamefont{Zheng}} \bibnamefont{and}
  \bibinfo{author}{\bibfnamefont{L.~K.} \bibnamefont{Wagner}},
  \bibinfo{journal}{Phys. Rev. Lett.} \textbf{\bibinfo{volume}{114}},
  \bibinfo{pages}{176401} (\bibinfo{year}{2015}).

\bibitem[{\citenamefont{Newton et~al.}(2014)\citenamefont{Newton, Sao,
  Fujisawa, Onitsuka, Kawaguchi, Tokuda, Sato, Togashi, Yabashi, Ishikawa
  et~al.}}]{Newton}
\bibinfo{author}{\bibfnamefont{M.~C.} \bibnamefont{Newton}},
  \bibinfo{author}{\bibfnamefont{M.}~\bibnamefont{Sao}},
  \bibinfo{author}{\bibfnamefont{Y.}~\bibnamefont{Fujisawa}},
  \bibinfo{author}{\bibfnamefont{R.}~\bibnamefont{Onitsuka}},
  \bibinfo{author}{\bibfnamefont{T.}~\bibnamefont{Kawaguchi}},
  \bibinfo{author}{\bibfnamefont{K.}~\bibnamefont{Tokuda}},
  \bibinfo{author}{\bibfnamefont{T.}~\bibnamefont{Sato}},
  \bibinfo{author}{\bibfnamefont{T.}~\bibnamefont{Togashi}},
  \bibinfo{author}{\bibfnamefont{M.}~\bibnamefont{Yabashi}},
  \bibinfo{author}{\bibfnamefont{T.}~\bibnamefont{Ishikawa}},
  \bibnamefont{et~al.}, \bibinfo{journal}{Nano lett.}
  \textbf{\bibinfo{volume}{14}}, \bibinfo{pages}{2413} (\bibinfo{year}{2014}).

\bibitem[{\citenamefont{Belozerov et~al.}(2012)\citenamefont{Belozerov,
  Korotin, Anisimov, and Poteryaev}}]{Belozerov}
\bibinfo{author}{\bibfnamefont{A.~S.} \bibnamefont{Belozerov}},
  \bibinfo{author}{\bibfnamefont{M.~A.} \bibnamefont{Korotin}},
  \bibinfo{author}{\bibfnamefont{V.~I.} \bibnamefont{Anisimov}},
  \bibnamefont{and} \bibinfo{author}{\bibfnamefont{A.~I.}
  \bibnamefont{Poteryaev}}, \bibinfo{journal}{Phys. Rev. B}
  \textbf{\bibinfo{volume}{85}}, \bibinfo{pages}{045109}
  (\bibinfo{year}{2012}).

\bibitem[{\citenamefont{Weber et~al.}(2012)\citenamefont{Weber, O'Regan, Hine,
  Payne, Kotliar, and Littlewood}}]{Weber}
\bibinfo{author}{\bibfnamefont{C.}~\bibnamefont{Weber}},
  \bibinfo{author}{\bibfnamefont{D.~D.} \bibnamefont{O'Regan}},
  \bibinfo{author}{\bibfnamefont{N.~D.~M.} \bibnamefont{Hine}},
  \bibinfo{author}{\bibfnamefont{M.~C.} \bibnamefont{Payne}},
  \bibinfo{author}{\bibfnamefont{G.}~\bibnamefont{Kotliar}}, \bibnamefont{and}
  \bibinfo{author}{\bibfnamefont{P.~B.} \bibnamefont{Littlewood}},
  \bibinfo{journal}{Phys. Rev. Lett.} \textbf{\bibinfo{volume}{108}},
  \bibinfo{pages}{256402} (\bibinfo{year}{2012}).

\bibitem[{\citenamefont{Kim et~al.}(2013)\citenamefont{Kim, Kim, Kang, and
  Min}}]{Kim}
\bibinfo{author}{\bibfnamefont{S.}~\bibnamefont{Kim}},
  \bibinfo{author}{\bibfnamefont{K.}~\bibnamefont{Kim}},
  \bibinfo{author}{\bibfnamefont{C.-J.} \bibnamefont{Kang}}, \bibnamefont{and}
  \bibinfo{author}{\bibfnamefont{B.~I.} \bibnamefont{Min}},
  \bibinfo{journal}{Phys. Rev. B} \textbf{\bibinfo{volume}{87}},
  \bibinfo{pages}{195106} (\bibinfo{year}{2013}).

\bibitem[{\citenamefont{Wall et~al.}(2012)\citenamefont{Wall, Wegkamp, Foglia,
  Appavoo, Nag, Jr., St\"{a}hler, and Wolf}}]{Wall}
\bibinfo{author}{\bibfnamefont{S.}~\bibnamefont{Wall}},
  \bibinfo{author}{\bibfnamefont{D.}~\bibnamefont{Wegkamp}},
  \bibinfo{author}{\bibfnamefont{L.}~\bibnamefont{Foglia}},
  \bibinfo{author}{\bibfnamefont{K.}~\bibnamefont{Appavoo}},
  \bibinfo{author}{\bibfnamefont{J.}~\bibnamefont{Nag}},
  \bibinfo{author}{\bibfnamefont{R.~F.~H.} \bibnamefont{Jr.}},
  \bibinfo{author}{\bibfnamefont{J.}~\bibnamefont{St\"{a}hler}},
  \bibnamefont{and} \bibinfo{author}{\bibfnamefont{M.}~\bibnamefont{Wolf}},
  \bibinfo{journal}{Nat. Commun.} \textbf{\bibinfo{volume}{3}},
  \bibinfo{pages}{721} (\bibinfo{year}{2012}).

\bibitem[{\citenamefont{Wall et~al.}(2013)\citenamefont{Wall, Foglia, Wegkamp,
  Appavoo, Nag, R.~F.~Haglund, St\"{a}hler, and Wolf}}]{Wall-2}
\bibinfo{author}{\bibfnamefont{S.}~\bibnamefont{Wall}},
  \bibinfo{author}{\bibfnamefont{L.}~\bibnamefont{Foglia}},
  \bibinfo{author}{\bibfnamefont{D.}~\bibnamefont{Wegkamp}},
  \bibinfo{author}{\bibfnamefont{K.}~\bibnamefont{Appavoo}},
  \bibinfo{author}{\bibfnamefont{J.}~\bibnamefont{Nag}},
  \bibinfo{author}{\bibfnamefont{J.}~\bibnamefont{R.~F.~Haglund}},
  \bibinfo{author}{\bibfnamefont{J.}~\bibnamefont{St\"{a}hler}},
  \bibnamefont{and} \bibinfo{author}{\bibfnamefont{M.}~\bibnamefont{Wolf}},
  \bibinfo{journal}{Phys. Rev. Lett.} \textbf{\bibinfo{volume}{87}},
  \bibinfo{pages}{115126} (\bibinfo{year}{2013}).

\bibitem[{\citenamefont{Momma and Izumi}(2011)}]{VESTA}
\bibinfo{author}{\bibfnamefont{K.}~\bibnamefont{Momma}} \bibnamefont{and}
  \bibinfo{author}{\bibfnamefont{F.}~\bibnamefont{Izumi}}, \bibinfo{journal}{J.
  Appl. Cryst.} \textbf{\bibinfo{volume}{44}}, \bibinfo{pages}{1272}
  (\bibinfo{year}{2011}).

\bibitem[{\citenamefont{Budai et~al.}(2014)\citenamefont{Budai, Hong, Manley,
  Specht, Li, Tischler, Abernathy, Said, Leu, Boatner et~al.}}]{Budai}
\bibinfo{author}{\bibfnamefont{J.~D.} \bibnamefont{Budai}},
  \bibinfo{author}{\bibfnamefont{J.}~\bibnamefont{Hong}},
  \bibinfo{author}{\bibfnamefont{M.~E.} \bibnamefont{Manley}},
  \bibinfo{author}{\bibfnamefont{E.~D.} \bibnamefont{Specht}},
  \bibinfo{author}{\bibfnamefont{C.~W.} \bibnamefont{Li}},
  \bibinfo{author}{\bibfnamefont{J.~Z.} \bibnamefont{Tischler}},
  \bibinfo{author}{\bibfnamefont{D.~L.} \bibnamefont{Abernathy}},
  \bibinfo{author}{\bibfnamefont{A.~H.} \bibnamefont{Said}},
  \bibinfo{author}{\bibfnamefont{B.~M.} \bibnamefont{Leu}},
  \bibinfo{author}{\bibfnamefont{L.~A.} \bibnamefont{Boatner}},
  \bibnamefont{et~al.}, \bibinfo{journal}{Nature}
  \textbf{\bibinfo{volume}{515}}, \bibinfo{pages}{535} (\bibinfo{year}{2014}).

\bibitem[{\citenamefont{Anisimov et~al.}(1991)\citenamefont{Anisimov, Zaanen,
  and Andersen}}]{Anisimov-LDA+U}
\bibinfo{author}{\bibfnamefont{V.~I.} \bibnamefont{Anisimov}},
  \bibinfo{author}{\bibfnamefont{J.}~\bibnamefont{Zaanen}}, \bibnamefont{and}
  \bibinfo{author}{\bibfnamefont{O.~K.} \bibnamefont{Andersen}},
  \bibinfo{journal}{Phys. Rev. B} \textbf{\bibinfo{volume}{44}},
  \bibinfo{pages}{943} (\bibinfo{year}{1991}).

\bibitem[{\citenamefont{Liechtenstein et~al.}(1995)\citenamefont{Liechtenstein,
  Anisimov, and Zaanen}}]{Liechtenstein}
\bibinfo{author}{\bibfnamefont{A.~I.} \bibnamefont{Liechtenstein}},
  \bibinfo{author}{\bibfnamefont{V.~I.} \bibnamefont{Anisimov}},
  \bibnamefont{and} \bibinfo{author}{\bibfnamefont{J.}~\bibnamefont{Zaanen}},
  \bibinfo{journal}{Phys. Rev. B} \textbf{\bibinfo{volume}{52}},
  \bibinfo{pages}{R5467} (\bibinfo{year}{1995}).

\bibitem[{\citenamefont{Gervais and Kress}(1985)}]{Gervais}
\bibinfo{author}{\bibfnamefont{F.}~\bibnamefont{Gervais}} \bibnamefont{and}
  \bibinfo{author}{\bibfnamefont{W.}~\bibnamefont{Kress}},
  \bibinfo{journal}{Phys. Rev. B} \textbf{\bibinfo{volume}{31}},
  \bibinfo{pages}{4809} (\bibinfo{year}{1985}).

\bibitem[{\citenamefont{McWhan et~al.}(1974)\citenamefont{McWhan, Marezio,
  Remeika, and Dernier}}]{McWhan}
\bibinfo{author}{\bibfnamefont{D.~B.} \bibnamefont{McWhan}},
  \bibinfo{author}{\bibfnamefont{M.}~\bibnamefont{Marezio}},
  \bibinfo{author}{\bibfnamefont{J.~P.} \bibnamefont{Remeika}},
  \bibnamefont{and} \bibinfo{author}{\bibfnamefont{P.~D.}
  \bibnamefont{Dernier}}, \bibinfo{journal}{Phys. Rev. B}
  \textbf{\bibinfo{volume}{10}}, \bibinfo{pages}{490} (\bibinfo{year}{1974}).

\bibitem[{\citenamefont{Kresse and
  Furthm\"{u}ller}(1996{\natexlab{a}})}]{VASP-1}
\bibinfo{author}{\bibfnamefont{G.}~\bibnamefont{Kresse}} \bibnamefont{and}
  \bibinfo{author}{\bibfnamefont{J.}~\bibnamefont{Furthm\"{u}ller}},
  \bibinfo{journal}{Phys. Rev. B} \textbf{\bibinfo{volume}{54}},
  \bibinfo{pages}{11169} (\bibinfo{year}{1996}{\natexlab{a}}).

\bibitem[{\citenamefont{Kresse and
  Furthm\"{u}ller}(1996{\natexlab{b}})}]{VASP-2}
\bibinfo{author}{\bibfnamefont{G.}~\bibnamefont{Kresse}} \bibnamefont{and}
  \bibinfo{author}{\bibfnamefont{J.}~\bibnamefont{Furthm\"{u}ller}},
  \bibinfo{journal}{Comput. Mat. Sci.} \textbf{\bibinfo{volume}{6}},
  \bibinfo{pages}{15} (\bibinfo{year}{1996}{\natexlab{b}}).

\bibitem[{\citenamefont{Kresse and Hafner}(1993)}]{VASP-3}
\bibinfo{author}{\bibfnamefont{G.}~\bibnamefont{Kresse}} \bibnamefont{and}
  \bibinfo{author}{\bibfnamefont{J.}~\bibnamefont{Hafner}},
  \bibinfo{journal}{Phys. Rev. B} \textbf{\bibinfo{volume}{47}},
  \bibinfo{pages}{558} (\bibinfo{year}{1993}).

\bibitem[{\citenamefont{Kresse and Joubert}(1999)}]{VASP-PAW}
\bibinfo{author}{\bibfnamefont{G.}~\bibnamefont{Kresse}} \bibnamefont{and}
  \bibinfo{author}{\bibfnamefont{D.}~\bibnamefont{Joubert}},
  \bibinfo{journal}{Phys. Rev. B} \textbf{\bibinfo{volume}{59}},
  \bibinfo{pages}{1758} (\bibinfo{year}{1999}).

\bibitem[{\citenamefont{Goodman}(1995)}]{Goodman}
\bibinfo{author}{\bibfnamefont{J.}~\bibnamefont{Goodman}},
  \emph{\bibinfo{title}{{Introduction to Fourier optics}}}
  (\bibinfo{publisher}{McGraw-Hill}, \bibinfo{address}{New York, NY},
  \bibinfo{year}{1995}).

\bibitem[{\citenamefont{Longo and Kierkegaard}(1970)}]{Longo}
\bibinfo{author}{\bibfnamefont{J.~M.} \bibnamefont{Longo}} \bibnamefont{and}
  \bibinfo{author}{\bibfnamefont{P.~A.} \bibnamefont{Kierkegaard}},
  \bibinfo{journal}{Acta Chem. Scand.} \textbf{\bibinfo{volume}{24}},
  \bibinfo{pages}{420} (\bibinfo{year}{1970}).

\bibitem[{\citenamefont{von Laue}(1936)}]{Laue1936}
\bibinfo{author}{\bibfnamefont{M.}~\bibnamefont{von Laue}},
  \bibinfo{journal}{Annalen der Physik} \textbf{\bibinfo{volume}{26}},
  \bibinfo{pages}{55} (\bibinfo{year}{1936}).

\bibitem[{\citenamefont{Miao et~al.}(2000)\citenamefont{Miao, Kirz, and
  Sayre}}]{Miao2000}
\bibinfo{author}{\bibfnamefont{J.}~\bibnamefont{Miao}},
  \bibinfo{author}{\bibfnamefont{J.}~\bibnamefont{Kirz}}, \bibnamefont{and}
  \bibinfo{author}{\bibfnamefont{D.}~\bibnamefont{Sayre}},
  \bibinfo{journal}{Acta Crystallographica Section D}
  \textbf{\bibinfo{volume}{56}}, \bibinfo{pages}{1312} (\bibinfo{year}{2000}).

\bibitem[{\citenamefont{Newton et~al.}(2010)\citenamefont{Newton, Leake,
  Harder, and Robinson}}]{Newton2010}
\bibinfo{author}{\bibfnamefont{M.~C.} \bibnamefont{Newton}},
  \bibinfo{author}{\bibfnamefont{S.~J.} \bibnamefont{Leake}},
  \bibinfo{author}{\bibfnamefont{R.}~\bibnamefont{Harder}}, \bibnamefont{and}
  \bibinfo{author}{\bibfnamefont{I.~K.} \bibnamefont{Robinson}},
  \bibinfo{journal}{Nature Materials} \textbf{\bibinfo{volume}{9}},
  \bibinfo{pages}{120} (\bibinfo{year}{2010}), ISSN \bibinfo{issn}{1476-1122}.

\bibitem[{\citenamefont{Robinson and Miao}(2004)}]{Robinson2004}
\bibinfo{author}{\bibfnamefont{I.}~\bibnamefont{Robinson}} \bibnamefont{and}
  \bibinfo{author}{\bibfnamefont{J.}~\bibnamefont{Miao}}, \bibinfo{journal}{MRS
  Bulletin} \textbf{\bibinfo{volume}{29}}, \bibinfo{pages}{177}
  (\bibinfo{year}{2004}), ISSN \bibinfo{issn}{0883-7694}.

\bibitem[{\citenamefont{Robinson and Harder}(2009)}]{Robinson2009}
\bibinfo{author}{\bibfnamefont{I.}~\bibnamefont{Robinson}} \bibnamefont{and}
  \bibinfo{author}{\bibfnamefont{R.}~\bibnamefont{Harder}},
  \bibinfo{journal}{Nature Materials} \textbf{\bibinfo{volume}{8}},
  \bibinfo{pages}{291} (\bibinfo{year}{2009}), ISSN \bibinfo{issn}{1476-1122}.

\end{thebibliography}

\newpage
\begin{center}
\Large{\textbf{Supplementary Material - \textit{Ab initio} molecular dynamics study of the structural and electronic transition in VO$_2$}}
\end{center}

\section{Effect of Hubbard parameters on structural, energetic and electronic properties of VO$_2$}

In our simulations, we used three different settings for U and J parameters,
which yield different lattice parameters, M1 band gap and energy difference
$\Delta$E between monoclinic M1 and rutile R phases at T = 0.
These different settings of the parameters also resulted in different
T$_c$(U,J) observed in the MD simulations and also in different energy
difference $\Delta$E at the corresponding T = T$_c$(U,J).
Results are summarized in Tables I and II.

\begin{center}
\begin{table}[hb]
\begin{tabular}{|c|c|c|c|c|}
\hline

Hubbard parameters [eV] & PBE (U=J=0) & U = 2.4, J = 0.5 & U = 3.15, J = 0.6 & U = 4.2, J = 0.8 \\ \hline

\textbf{M1} \\ \hline
$a$ [\AA]		& 5.634  & 5.660  & 5.666  &  5.681    \\ \hline
$b$ [\AA] 		& 4.560  & 4.601  & 4.605  & 4.607     \\ \hline
$c$ [\AA] 		& 5.413  & 5.437  &  5.443 & 5.449     \\ \hline
$\beta$ [$\degree$]     & 121.88 & 122.03 & 122.04 &  122.10   \\ \hline
M1 band gap [eV]            &  0.0   &  0.25  &  0.36  &   0.56   \\ \hline

\textbf{R} \\ \hline
$a$ [\AA]		& 4.616 & 4.626 & 4.631 & 4.637 \\ \hline
$c$ [\AA]		& 2.773 & 2.792 & 2.797 & 2.799 \\ \hline

\end{tabular}
\caption{Lattice parameters of M1 and R and M1 band gap obtained for different U and J parameters.}
\end{table}
\end{center}

\begin{center}
\begin{table}[hb]
\begin{tabular}{|c|c|c|c|c|}
\hline

Hubbard parameters [eV] & PBE (U=J=0) & U = 2.4, J = 0.5 & U = 3.15, J = 0.6 & U = 4.2, J = 0.8 \\ \hline \hline
MD-estimated T$_c$ [K]                   &        &  300   &   450  &   750          \\ \hline \hline
$\Delta$E(T$_c$) [meV/atom] &        &  5     &   11   &   15           \\ \hline \hline
$\Delta$E(T=0)   [meV/atom] & -3.7   &  12.2  &   21.9 &   35.0           \\ \hline

\end{tabular}
\caption{MD-obtained T$_c$ and energy differences $\Delta$E at T = 0 and at T=T$_c$(U,J) for different U and J.
Experimental latent heat at T$_c$ = 340 K was measured to be 14.7 meV/atom \cite{Budai}.}
\end{table}
\end{center}

The optimal U, J setting that yields best compromise between the band gap,
energetics of VO$_2$ phases and MD-estimated T$_c$ between M1 and R was found
to be U = 3.15 eV, J = 0.6 eV with T$_c$=450 K. The corresponding band structure
of M1 and R phases and projected densities of states (peDOS) calculated
with these parameters are shown in Figs. S1 and S2, respectively.

\begin{figure}[h]
\begin{tabular}{c|c}
\includegraphics[width=0.45\columnwidth]{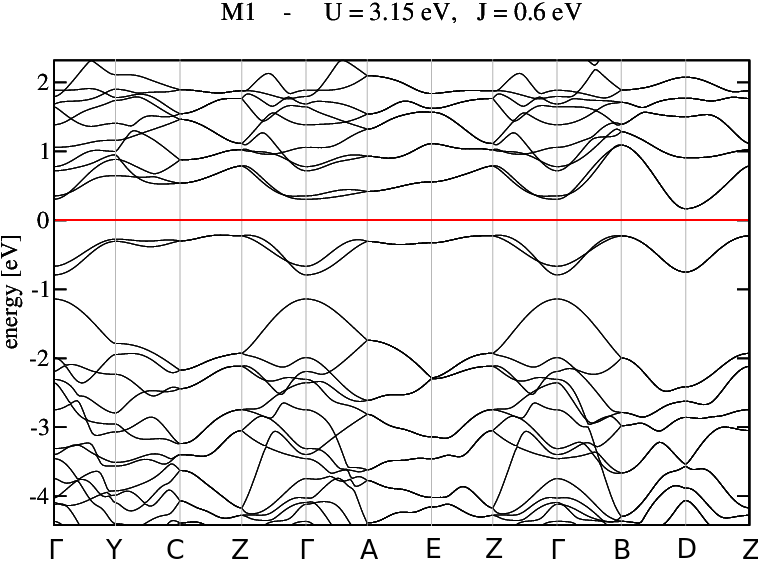}
\includegraphics[width=0.45\columnwidth]{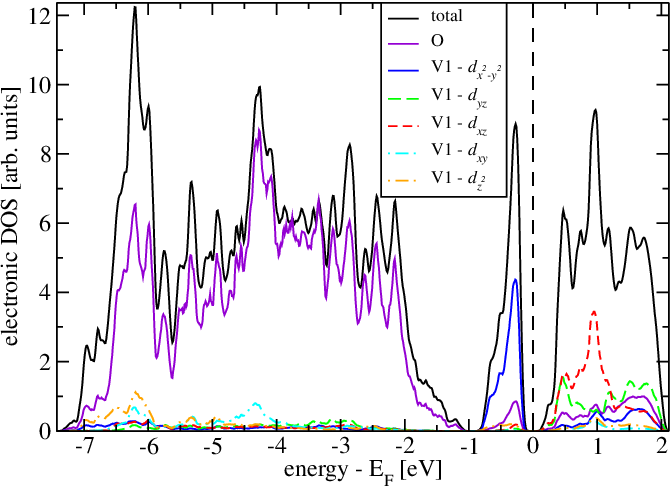}
\end{tabular}
\caption*{Fig.~S1. Band structure along $\Gamma$-Y-C-Z-$\Gamma$-A-E-Z-$\Gamma$-B-D-Z path
and peDOS in the local geometry of V1 atoms \cite{Eyert-review} for optimized M1 phase.}
\end{figure}

\begin{figure}
\begin{tabular}{c|c}
\includegraphics[width=0.45\columnwidth]{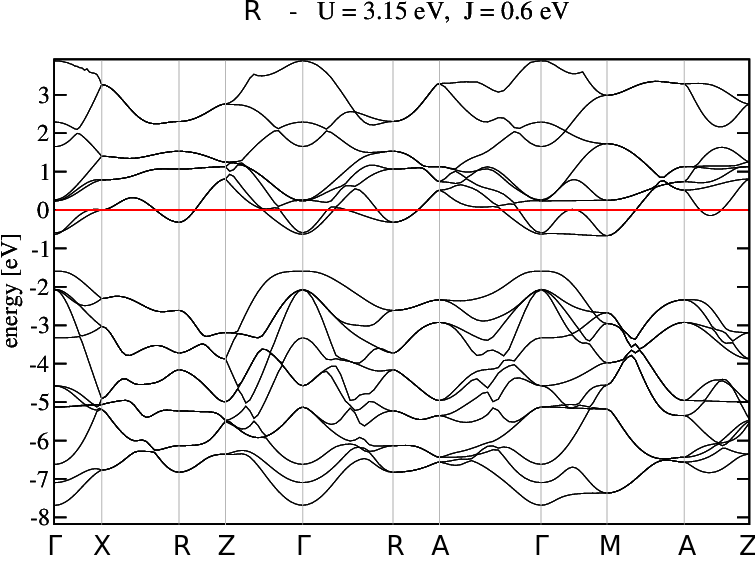}
\includegraphics[width=0.45\columnwidth]{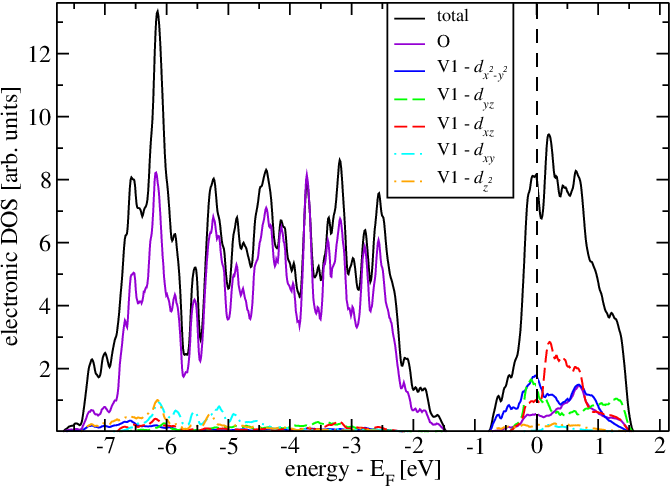}
\end{tabular}
\caption*{Fig.~S2. Band structure along $\Gamma$-X-R-Z-$\Gamma$-R-A-$\Gamma$-M-A-Z path and peDOS in the local
geometry of V1 atoms \cite{Eyert-review} for optimized R phase.}
\end{figure}

\section{Definition of order parameters}

The actual structural state of the system during the MD simulations was determined
by tracking the dimerization amplitudes (DA) of individual chains of V atoms and
the average tilting parameter of V-V dimers. We define the dimerization
amplitude of a chain by the formula

\begin{center}
$d = \frac{1}{N} \sum^n_{j=1} (-1)^j \Delta x_j$,
\end{center}

where $j = 1 ... n$ is the index of a chain atom (with total of $n$ atoms in
a chain within the supercell) and $\Delta x_j$ is the deviation of V atom coordinate
from its ideal position in the R phase. The dimerization amplitude for each $i$-th chain - $d_i$
hence for the R phase equals zero, while for M1 it is 0.169 \AA \,for an optimized (T=0)
structure. There are 32 chains in our simulation sample, each with 8 V atoms,
whose DA were averaged to obtained mean DA.

The tilting order parameter is defined here as the average angle $\delta$ between
the dimers and the $\vec{a}$-axis (in M1 supercell geometry). The twisting angle
yields 0 in the R phase and $\approx$ 7.5$\degree$ in T=0 M1. There are 128 dimers
in our simulation sample and the tilting parameter was taken as the average twisting
angle from all these dimers. 

\section{Evolution at 400 K before the M1-R transition}

As described in the main text, at T=400 K the system simulated with U=3.15 eV and
J=0.6 eV exhibited large fluctuations of DAs towards low values - Fig.~S3, which corresponds
to the fact that some chains were fairly disrupted during some time intervals within
the MD run. This regular loss of the long-range order between some dimers is an indication
that the transitions from M1 to R already tries to initiate.

\begin{figure}
\includegraphics[width=0.7\columnwidth]{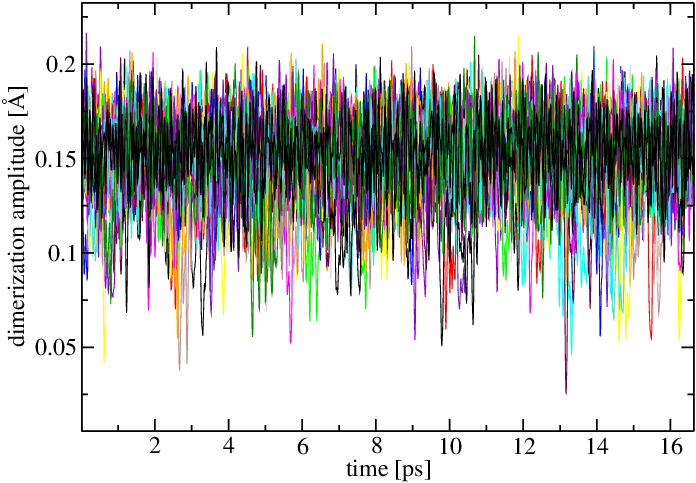}
\caption*{Fig.~S3. Evolution of dimerization amplitudes at 400 K calculated with
U=3.15 eV and J=0.6 eV showing large fluctuations of certain chains in some time intervals.}
\end{figure}

After rising the temperature to 450 K, the M1 phase quickly transformed into R, as described
in the main text. The pictures of the actual simulation supercells with 768 atoms before and after
this process are shown in Fig.~S4.

\begin{figure}
\includegraphics[width=0.99\columnwidth]{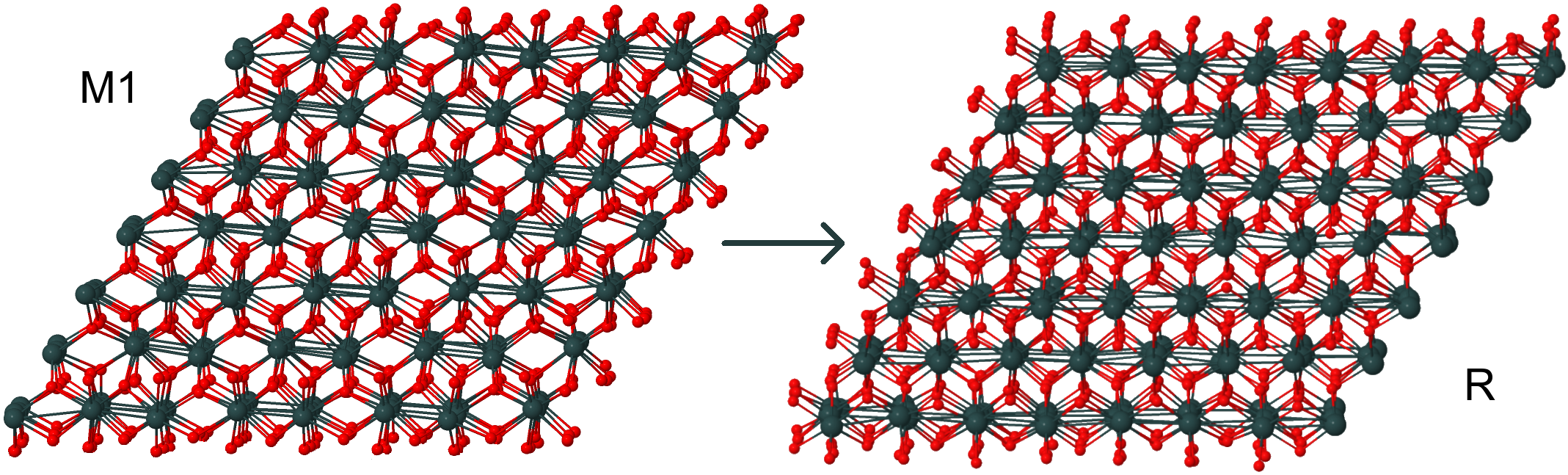}
\caption*{Fig.~S4. 768-atomic supercells used in the simulations - M1 before the transition
at 400 K and R phase at 450 K.}
\end{figure}

\section{Transitions at different temperatures}

The M1-R transformations were observed at different temperatures when different U, J
settings were used and the main results are summarized in Table II. The corresponding
evolutions of DA are shown in Fig.~S5 for $T_c=300$ K (U=2.4 eV, J=0.5) and for $T_c=750$ K
(U=4.2 eV, J=0.8 eV).

\begin{figure}
\begin{tabular}{c|c}
\includegraphics[width=0.45\columnwidth]{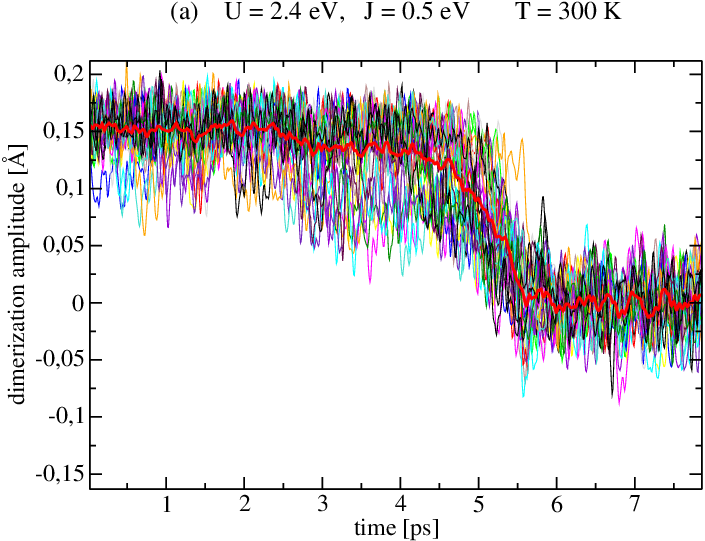}
\includegraphics[width=0.45\columnwidth]{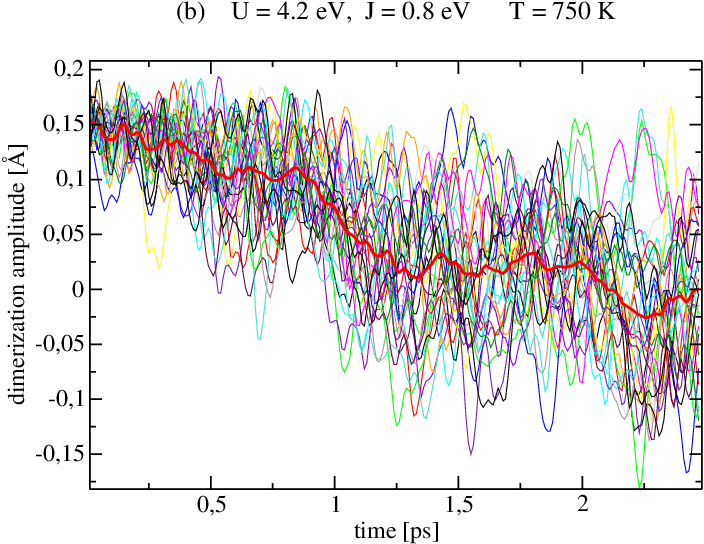}
\end{tabular}
\caption*{Fig.~S5. Evolution of dimerization amplitudes at 300 K (a) and at 750 K (b).}
\end{figure}

\section{Atomic fluctuations in M1 and R phases}

The calculated values of static fluctuations of V and O atoms
(calculated with U=3.15 eV and J=0.6 eV) show good agreement
with experimental findings - Table \ref{fluctuations}.
Fluctuations of lighter O atoms are greater than fluctuations of V atoms in
the M1 phase at 300 K by a factor of 1.4, while in R phase at 450 K the
opposite is true and $\left<u\right>^2_V$ is nearly 1.5 times higher than
$\left<u\right>^2_O$, according to the current simulations.

\begin{center}
\begin{table}[h]
\begin{tabular}{|c|c|c|c|}
\hline
fluctuations [\AA$^2$] & experiment - Ref. \cite{Longo} & simulations - Ref. \cite{Budai} & current simulations \\ \hline
\textbf{M1}   \\ \hline
 & T = 298 K &   & T = 300 K \\ \hline
$\left<u^2\right>_V$  &  0.0114  &   &  0.0102 \\ \hline
$\left<u^2\right>_O$  &  0.0159  &   &  0.0142  \\ \hline
\textbf{R}   \\ \hline
 & T = 470 K & T = 425 K & T = 450 K \\ \hline
$\left<u^2\right>_V$  &  0.036  &  0.037 &  0.0348  \\ \hline 
$\left<u^2\right>_O$  &  0.028  &  0.021 &  0.0235  \\ \hline 
\end{tabular}
\caption{Calculated and experimental values of fluctuations of V and O atoms in M1 and R phases. Current simulations are given in the third column.}
\label{fluctuations}
\end{table}
\end{center}

These large fluctuations of V atoms in the R phase are the well-known cause
for the large Debye-Waller factors that are commonly observed in XRD experiments \cite{McWhan}.
This implies that V atoms remain relatively long time far away from their mean
positions in R indicating a flat energy surface for the variation of V atomic positions,
in contrast to M1, where the movement of V atoms in dimers is much more constrained.
This might possibly be attributed to the strong tendency of vanadium atoms to
dimerize even at the P-T conditions corresponding to the stable R phase region.

\section{Coherent X-ray Diffraction during the M1-R transition}

Bragg coherent X-ray diffraction imaging (BCXDI) is a lens-less imaging technique that permits imaging of crystalline materials with a sub-angstrom sensitivity.  It is largely non-destructive and provides strain tensor information at the surface and throughout the bulk of a material. Experimentally, BCXDI is routinely performed at 3\textsuperscript{rd} generation synchrotron facilities by illuminating a sample with a spatially coherent X-ray source while ensuring that the coherence length exceeds the dimensions of the crystal. In the Bragg reflection geometry, scattered light from the crystal interferes in the far-field, producing a three-dimensional k-space speckle pattern. The diffracted intensity is measured using an area X-ray detector which is positioned far enough away from the sample to resolve the finest fringes of the speckle pattern. The third dimension is obtained by rotating the Ewald sphere through the Bragg condition while maintaining a largely fixed incident ($\mathbf{k}_i$) and reflected ($\mathbf{k}_f$) wave vector. Iterative phase reconstruction methods are then used to recover the complex three-dimensional electron density and phase information. The displacement of ions throughout the bulk is directly related to the phase and can be used to obtain strain information according to the relation  $\phi = \mathbf{Q} \cdot \mathbf{u}$, where $\mathbf{Q}$ is a particular reciprocal lattice point and $\mathbf{u}$ is the atomic displacement.\cite{Laue1936, Miao2000, Newton2010, Robinson2004, Robinson2009}

With the advent of 4\textsuperscript{th} generation X-ray free electron laser (XFEL) synchrotron facilities, it has become possible to study ultra-fast structural dynamics using femto-second coherent X-ray pulses. When combined with a femto-second optical excitation source, it is possible to perform stroboscopic measurements in a pump-probe scheme.\cite{Newton} 

In order to contrast \textit{ab initio} calculations with ultra-fast pump-probe coherent X-ray diffraction experiments, we can define a static reference lattice at the initial low temperature phase of the system (i.e the M1 phase).  This is achieved by averaging thermal fluctuations of each atomic position over a predefined time interval ($\Delta t$) such that:
\begin{equation}
 \mathbf{p}_j = \frac{1}{\Delta t}\sum _{t_0} \mathbf{r}_j(t) .
\end{equation}
For each subsequent time frame, the displacement of each atom from this equilibrium position is then given by:
\begin{equation}
\mathbf{u}_j(t) = \mathbf{r}_j(t) - \mathbf{p}_j .
\end{equation}
The spatially varying far-field scattering amplitude distribution is proportional to the exponentiated phase associated with scattering from each atom in the supercell:
\begin{equation}
A(\mathbf{q},t) \propto \sum _j f_j \; e^{-i\mathbf{q} \cdot \mathbf{u}_j(t)}
\end{equation}
where $\mathbf{q} = \mathbf{k} - \mathbf{k}_i$ is the wavevector transfer between the incident wavevector $\mathbf{k}_i$ and a general reflected scattering vector $\mathbf{k}$ and $f_j$ is the atomic scattering factor.  

We monitored changes in the intensity $I(\mathbf{q},t) = |A (\mathbf{q},t)|^2$ from the speckle pattern that results from the (011) Bragg reflection in order to obtain a direct comparison with ultra-fast femto-second coherent X-ray diffraction experiments performed on nanoscale crystals.

\end{document}